\documentclass{article}

\usepackage{arxiv}

\usepackage[utf8]{inputenc} 
\usepackage[T1]{fontenc}    
\usepackage{hyperref}       
\usepackage{url}            
\usepackage{booktabs}       
\usepackage{amsfonts}       
\usepackage{nicefrac}       
\usepackage{microtype}      
\usepackage{lipsum}		    
\usepackage{graphicx}
\usepackage{natbib}
\usepackage{doi}
\usepackage{array}
\usepackage{siunitx}
\usepackage{amsmath}
\usepackage{etoolbox}
\usepackage{float}

\renewrobustcmd{\bfseries}{\fontseries{b}\selectfont}
\renewrobustcmd{\boldmath}{}
\newrobustcmd{\B}{\bfseries}

\title{K-Anonymous A/B Testing}

\author{Matt Gershoff \\
    Conductrics INC \\
    Austin TX 78748\\
    \texttt{matt.gershoff@conductrics.com}
}

\date{}

\hypersetup{
pdftitle={Privacy Engineering for A/B Testing},
pdfauthor={Matt Gershoff},
}

\begin{document}
\maketitle

\begin{abstract}
	A core principle of Privacy by Design (PbD) is minimizing the data that is stored or shared about each individual respondent. PbD principles are mandated by the GDPR (see Article 5c and Article 25), as well as informing aspects of California Privacy Rights Act (CPRA). This paper describes a simple and effective approach that can be used in many a/b testing and similar contexts to help meet these PbD goals. Specifically, the method presented describes an approach to run OLS regression on k-anonymized data. To help illustrate the general utility of this approach, descriptions of two important use cases are offered: 1) calculating partial f-tests as a simple way to both check for a/b test interactions and to test for heterogeneity of treatment effects; and 2) regression adjustment using an approach similar to the popular CUPED method, as a variance reduction method for a/b tests. Using this method has advantages for privacy and compliance, as well as often reducing data storage and processing costs, by storing, sharing, or analyzing only aggregate level rather than individual level data.

\end{abstract}


\section{Introduction}
Principle 2 of PbD calls for and defines data minimization\citet{cavoukian2009}. 

\begin{quote}
"The collection of personally identifiable information should be kept to a strict minimum. The design of programs, information and communications technologies, and systems should begin with non-identifiable interactions and transactions, as the default. Where possible, the identifiability, observability, and linkability of personal information should be minimized."
\end{quote}

The principle of data minimization is explicitly outlined as part of GDPR requirements \citep{gdpr2016}. It turns out, happily, that many of the statistics needed for modern digital a/b testing programs, if we are intentional about how the data are collected or extracted, do not require access to identifiable, individual-level data. 

Most of the basic statistical approaches used for inference in a/b testing programs (ANOVA, ANCOVA, t-tests, partial f-tests, etc.) can be recast as different applications of Ordinary Least Squares (OLS) regression. What has been unappreciated is that OLS can be performed efficiently on data aggregated into equivalence classes. In addition, the use of equivalence classes also facilitates the use of k-anonymization as a data minimization approach consistent with principle 2 of PbD.  Although k-anonymity is unable to provide formal privacy guarantees, it does, however, offer a practical and useful measure of privacy and data minimization.

After a brief review of k-anonymization, the basics of performing OLS regression on data stored in equivalence classes (think pivot tables) are covered.  This paper will review two main regression designs: 

\begin{enumerate}
    \item Dummy variable design matrices.
    \item OLS with covariates that have cardinality strictly lower than the sample size.
\end{enumerate}

We will show that if following a regime of data minimization, the use of equivalence class data will yield the same regression results as when using unaggregated (user lever) data and simplify computing the OLS matrices ($X^\prime X$ and $X^\prime y$). 

For each case, we will walk through a working example to motivate and, hopefully, clarify the discussions. For dummy variable OLS we introduce the partial f-test method to test for both pairwise interactions between independent a/b tests and for heterogeneity of treatment effects. OLS with cardinal variables is illustrated with a very simple regression adjustment (CUPED) example.

\section{K-Anonymization and A/B Testing}

While the best approach to privacy will of course depend on the context and use case, to ensure that the privacy measure is simple to both manage and interpret, we propose the use of K-anonymity. While K-anonymity does not provide the types of formal guarantees of differential privacy, it does provide a very simple, yet useful signal that both encourages data minimization and makes data privacy audits much simpler to perform and interpret. 

K-anonymity is based on the minimum number of individuals that have been aggregated into a smaller set of mutually exclusive classes \citep{samarati1998}. A larger k can be said to provide more protection against identification, as there are at least k-1 other individuals who are indistinguishable from any other individual within a dataset.  K-anonymity tends to increase in any given data structure when we reduce: 

\begin{enumerate}
\item The cardinality of the set of variables that have been linked.
\item The resolution (granularity) of the elements contained in each variable.
\end{enumerate}

These two properties, fewer data linkages and reduced data granularity, are of particular interest as they explicitly called for in principle 2 of privacy by design, and hence the GDPR. 

A set of equivalence classes is formed by enumerating over each of the unique configurations of the set of quasi-identifiers. A data field is considered a quasi-identifier if combined with other pieces of information, it has the potential to uniquely identify an individual.  The number of the classes will be upper bounded by the joint cardinality of the unique configuration of elements across the combination of the quasi-identifiers.  

For example, Table~\ref{tab:table1} is user level data containing:
\begin{enumerate}
\item The treatment a user was exposed to (‘A’ or ‘B’).
\item An associated covariate (‘1’, ‘2’, ‘3’).
\item Post treatment Time on App (e.g. the end point for an a/b test).
\end{enumerate}

\begin{table}[ht]
\centering
\caption{Time on App Data}
\label{tab:table1}
\begin{tabular}{
    l
    c
    c
    S[table-format=1.9]
}
\toprule
\textbf{User ID} & \textbf{Treatment} & \textbf{Covariate} & \textbf{Time on App } \\
\midrule
XXX1  & A & 1 & 1.035051833 \\
XXX2  & A & 2 & 2.052761778 \\
XXX3  & A & 3 & 0.818065309 \\
XXX4  & A & 1 & 0.445755940 \\
XXX5  & A & 2 & 2.192969747 \\
XXX6  & A & 3 & 1.351287088 \\
XXX7  & A & 1 & 0.690041547 \\
XXX8  & A & 2 & 1.965163288 \\
XXX9  & A & 3 & 0.885217757 \\
XXX10 & B & 1 & 0.836158231 \\
XXX11 & B & 2 & 0.224290470 \\
XXX12 & B & 3 & 2.689670150 \\
XXX13 & B & 1 & 0.286027328 \\
XXX14 & B & 2 & 0.668558807 \\
XXX15 & B & 3 & 2.138392457 \\
XXX16 & B & 1 & 0.300483978 \\
XXX17 & B & 2 & 0.816659567 \\
XXX18 & B & 3 & 2.406463861 \\
\bottomrule
\end{tabular}
\end{table}

Table~\ref{tab:table2} is an equivalence class representation of the data in Table~\ref{tab:table1}. Each ‘class’ is the unique combination of treatment assignment with the covariates. Each class stores a count the associated subjects with the class and the sum of the post treatment Time on App data for those subjects.

\begin{table}[ht]
\centering
\caption{Time on App Data Summary}
\label{tab:table2}
\label{tab:Time on App_summary}
\begin{tabular}{
    l
    l
    S[table-format=1]
    S[table-format=1]
    S[table-format=1.5]
}
\toprule
\multicolumn{1}{c}{} & \textbf{Treatment} & \textbf{Covariate} & \textbf{Count} & \textbf{Sum Time on App } \\
\midrule
R1 & A & 1 & 3 & 2.170849320 \\
R2 & A & 2 & 3 & 6.210894813 \\
R3 & A & 3 & 3 & 3.054570154 \\
R4 & B & 1 & 3 & 1.422669536 \\
R5 & B & 2 & 3 & 1.709508844 \\
R6 & B & 3 & 3 & 7.234526468 \\
\bottomrule
\end{tabular}
\end{table}

Notice that while the maximum cardinality of the individual level data in Table~\ref{tab:table1} is $N$ (the number of subjects), the maximum equivalence class data size, $M$, is 6, regardless of the sample size. $M$ is determined by the joint cardinality of the quasi-identifiers under consideration – $m$ equals six as there are two treatment options and three values of the covariate.  Of course, in general, as the cardinality of the linked variables and/or the fidelity of the variables increases, then $M$ can quickly grow to $N$. However, by following principle 2 we are intentionally trying to ensure that  $M \ll N$.

Table~\ref{tab:table3} contains the sums of squares for the Time on App data per test treatment arm.  While Table~\ref{tab:table2} provides enough information for unbiased estimation, for inference (t-test, f-tests etc.), we need the total sum of squares (TSS) for each end point, or goal associated with the experiment. In our example, this is the sum of squares of the Time on App, $\sum_{i=1}^N y_i^2$. To account for possible heteroskedastic errors across the treatments, a TSS is collected for each treatment arm ($\sum_{a \in A} y_{a}^2$, $\sum_{b \in B} y_{b}^2$).

\begin{table}[ht]
\centering
\caption{Sum of Squared Time on App by Treatment Group}
\label{tab:table3}
\begin{tabular}{
    l
    S[table-format=2.3]
}
\toprule
\textbf{Treatment} & \textbf{\(\sum \text{Time on App}^2\)} \\
\midrule
A & 17.910 \\
B & 19.634 \\
\bottomrule
\end{tabular}
\end{table}

Although one could calculate and add the sum of squares data to equivalence class in Table~\ref{tab:table2}, the finer the detail we store the sum of squares data, the more information about how the Time on App data is distributed is exposed.  L-diversity, an extension of k-anonymization, is a measure of how distributed or concentrated the endpoint attribute is \citep{machanavajjhala2007}.  If the attribute itself is sensitive data, we would like to ensure that diversity of values is high in order to reduce the probability that an attacker can learn this value for any individual within the class.   Since the sum of squares is a measure of the variance of the attribute, it provides additional information that can be used to help unpack how the summary Time on App data are distributed.  For example, the sum of squares for class ‘R1’ in Table 2 is 4.7089. An attacker would then know that within R1, two people did not spend any time, and one person spent 2.17 (since $\sqrt{4.7089}=2.17$). If the attacker knew that the person of interest was in treatment ‘A’ and had a value of ‘1’ for the covariate AND that they had spent any time on the app, they would be able to learn how much time they spent. Or conversely, if they knew that a person was in the data set, and had made a spent 2.17 then the attacker would learn that that person had been exposed to treatment ‘A’ and had a value of ‘1’ for the covariate. For this reason, it can be safer to calculate and store the sum of squares data only at the level of the required analysis. \footnote{One might think to minimize t-closeness – an extension of the k-anonymization framework that is a measure of how different the distribution of the sensitive end point is within each equivalence class versus the overall data set \citep{li2007}. However, this is not useful in our case because we want to use the differences in distributions to learn if there is a relationship between the covariates and the endpoint.}

\section{Local vs Global K-Anonymization}

\subsection{The Local Approach}

Borrowing from Differential Privacy concepts of local and global privacy, we will call a local k-anonymization approach one that only collects and stores data in aggregate form.  In the local k-annomized a/b testing case the test assignment is made on the client machine. The client then emits telemetry back to the server with the assignment and, if desired, any associated quasi-identifiers.  For example, a visitor is entered into Test 1 and assigned to treatment ‘B’ with a covariate value of ‘3’. The telemetry sent back to the server is ‘Test1:B:3’ which updates row R6 in Table~\ref{tab:table2} by increasing the Count field by 4 (3+1).  If the visitor then spends 4 time units on the app, the original message is sent as a key with the Time on App amount appended, ‘Test1:B:3:Time on App:4’. Row R6 in Table~\ref{tab:table2} is updated again so that the Sum Time on App field is increased by 4 to 11.23 and the total sum of squares for ‘B’ in Table~\ref{tab:table3} is incremented by 16 to 25.634. If the same user during another session spends 2 time units, the client sends Test1:B:Time on App\_old:4:Time on App:2 – and we increment the sum of squares by 20, rather than 4. Why? Because the total Time on App for this subject is 6. The square of 6 is 36. Since in the first Time on App message we added 16 to the sum of squares, we need to add 20 more, rather than naïvely adding only $2^2$. To just add an additional 4 would artificially reduce the calculated variance, inflating our p-values and biasing our test results.  

\subsection{The Global Approach}

In the global approach, the individual level data is collected and stored, but it is walled off and under the control of a secure curator. The curator can then release the appropriately aggregated data sets to internal and/or external teams. This approach can help ensure proper governance and control of potentially sensitive information. Rather than creating a multitude of copies and/or mirrors of sensitive data, which increases the odds of a privacy leaking release, only copies of aggregate data, which are of much lower risk, leaves the secure curator’s control.  One way to think of equivalence classes under the global approach is as a SQL select with the treatment and the quasi-identifiers in a group by statement.  As we shall see, this $\mathcal{O}(n)$ operation will let us convert what would have been a $\mathcal{O}(n^2 k)$ to a $\mathcal{O}(m^2 k)$ compute, with $m\ll n$ when applying the data minimization principle. 

\section{OLS for Categorical Covariates: Dummy Codes}
Many a/b testing program tasks can be reduced to OLS on dummy coded data. These tasks include:

\begin{enumerate}
    \item Estimates of average treatment effects (ATE) and their associated t-tests.
    \item Interference/interaction tests between independent a/b tests.
    \item Tests to check for possible conditional ATEs (heterogenous treatment effects) given categorical contextual or side information about the user. 
\end{enumerate}

\section{Equivalence Class to Gramian Matrices}
The standard normal equations for OLS and their covariance structures are
\[
\beta = (X^\prime X) ^{-1} X^\prime y \text{ and } \sigma^2 (X^\prime X)^{-1}.
\]

The Gramian matrix ($X^\prime X$) is constructed from the inner product of the design matrix ($X$), which is calculated as the sum of squares (along the main diagonal) and the sums of cross products (on the off diagonal) of the covariate data as follows:  

\[
X^\prime X =
\begin{bmatrix}
\sum_{1}^{N} x_1^2 & \cdots & \sum_{1}^{N} x_1 x_k \\
\vdots & \ddots & \vdots \\
\sum_{1}^{N} x_k x_1 & \cdots & \sum_{1}^{N} x_k^2
\end{bmatrix}
\]

The complexity of constructing the Gramian matrix from micro/user level data is $\mathcal{O}(n^2 k)$, where n is the number of subjects (users), and k is the number of covariates in the design matrix.  However, what is not often realized is that for categorical, dummy variable design matrices, the above matrix can be simplified to:  

\[
X^\prime X =
\begin{bmatrix}
\text{Count}(x_1) & \cdots & \text{Count}(x_1 = 1 \text{ and } x_k = 1) \\
\vdots & \ddots & \vdots \\
\text{Count}(x_k = 1 \text{ and } x_1 = 1) & \cdots & \text{Count}(x_k)
\end{bmatrix}
\]

where the Gramian matrix is composed only of simple and conditional counts. Fortunately, this count data is already mostly precomputed in the equivalence class format, leading to an extremely efficient construction of the Gramian matrix without the need to access, and perform more memory, or compute intensive queries (e.g SQL), on the microdata. 

Similarly, for the dummy coding case, the $X^\prime y$ matrix reduces to just conditional sums of $Y$, also readily available with very little compute resources from the equivalence class data.

\[
X^\prime y =
\begin{bmatrix}
\sum_{1}^{N} x_1 y \\
\sum_{1}^{N} x_2 y \\
\vdots \\
\sum_{1}^{N} x_k y
\end{bmatrix}
=
\begin{bmatrix}
\sum_{1}^{N} \big(y \mid x_1 = 1\big) \\
\sum_{1}^{N} \big(y \mid x_2 = 1\big) \\
\vdots \\
\sum_{1}^{N} \big(y \mid x_k = 1\big)
\end{bmatrix}
\]

\section{Example}

For example, Table~\ref{tab:table4} is a dummy encoding of the data in Table~\ref{tab:table1}, such that treatment ‘B’ and ‘2’ and ‘3’ of the covariate values are each dummy variables. 

\begin{table}[ht]
\centering
\caption{Dummy Encoding of Table 1 Data}
\label{tab:table4}
\begin{tabular}{cccc}
\toprule
\textbf{Int} & \textbf{Treatment} & \textbf{Covariate\_2} & \textbf{Covariate\_3} \\
\midrule
1 & 0 & 0 & 0 \\
1 & 0 & 0 & 1 \\
1 & 0 & 1 & 0 \\
1 & 0 & 0 & 1 \\
1 & 0 & 1 & 0 \\
1 & 0 & 0 & 1 \\
1 & 0 & 1 & 0 \\
1 & 0 & 0 & 1 \\
1 & 1 & 0 & 0 \\
1 & 1 & 0 & 1 \\
1 & 1 & 1 & 0 \\
1 & 1 & 0 & 1 \\
1 & 1 & 1 & 0 \\
1 & 1 & 0 & 1 \\
1 & 1 & 1 & 0 \\
1 & 1 & 0 & 1 \\
\bottomrule
\end{tabular}
\end{table}

The calculated Gramian matrix for Table~\ref{tab:table4} is 

\[
X^\prime X =
\begin{bmatrix}
18 & 9 & 6 & 6 \\
9 & 9 & 3 & 3 \\
6 & 3 & 6 & 0 \\
6 & 3 & 0 & 6
\end{bmatrix}
\]

which can be constructed using the standard matrix transpose and multiplication operations on the design matrix in Table~\ref{tab:table4}. However, the same Gramian matrix can also be quickly constructed from Table~\ref{tab:table2} by summing the count field from the appropriate rows as follows: 

\[
X^\prime  X =
\begin{bmatrix}
\textit{All Rows} & R4, R5, R6 & R2, R5 & R3, R6 \\
R4, R5, R6 & R4, R5, R6 & R5 & R6 \\
R2, R5 & R5 & R2, R5 & 0 \\
R3, R6 & R6 & 0 & R3, R6
\end{bmatrix}.
\]

For $X^\prime y$, we follow a similar pattern, but instead of selecting values from the count column, we add the rows from the Sum Time on App field as follows:

\[
X^\prime y =
\begin{bmatrix}
\textit{All Rows} \\
R4, R5, R6 \\
R2, R5 \\
R3, R6
\end{bmatrix}
=
\begin{bmatrix}
21.8030 \\
10.3667 \\
7.9204 \\
10.2891
\end{bmatrix}
\]

This yields the same result as if we had performed matrix operations on the micro-data in Table~\ref{tab:table4} appended with the Time data from Table~\ref{tab:table1}.  We can apply standard matrix operations on our two matrices to estimate the regression weights via the normal equations,  $\beta=(X^\prime X )^{-1} X^\prime y$.  For this example, we get the following result:

\[
\boldsymbol{\beta} =
\begin{bmatrix}
\beta_0 \\
\beta_{\text{treatment\_b}} \\
\beta_{\text{covariate\_2}} \\
\beta_{\text{covariate\_3}}
\end{bmatrix}
=
\begin{bmatrix}
18 & 9 & 6 & 6 \\
9 & 9 & 3 & 3 \\
6 & 3 & 6 & 0 \\
6 & 3 & 0 & 6
\end{bmatrix}^{-1}
\begin{bmatrix}
21.8030 \\
10.3667 \\
7.9204 \\
10.2891
\end{bmatrix}
=
\begin{bmatrix}
0.6583 \\
-0.1188 \\
0.7211 \\
1.1159
\end{bmatrix}
\]

\section{Standard OLS Errors}

Once we have estimated the regression weights, it is a simple, and efficient process to uncover the OLS error structure.   Recall that the residual sum of squares is  $e^\prime e$ = $\sum (y-X\boldsymbol{\beta})^2$ , which is the total sum of squares (TSS) minus the regression sum of squares. The TSS for the pooled regression from Table~\ref{tab:table3} is $TSS_{pooled} = 17.91 + 19.63 = 37.543$. The regression sum of squares can be computed using $\boldsymbol{\beta}^\prime (X^\prime X)\boldsymbol{\beta}$. Since we already have both $\boldsymbol{\beta}$ and $X^\prime X$ we can quickly compute the regression sum of squares. In the above example this resolves to

\[
\boldsymbol{\beta}^\prime X^\prime X \boldsymbol{\beta} =
\begin{bmatrix}
0.6583 & -0.1188 & 0.7211 & 1.1159
\end{bmatrix}
\begin{bmatrix}
18 & 9 & 6 & 6 \\
9 & 9 & 3 & 3 \\
6 & 3 & 6 & 0 \\
6 & 3 & 0 & 6
\end{bmatrix}
\begin{bmatrix}
0.6583 \\
-0.1188 \\
0.7211 \\
1.1159
\end{bmatrix}
= 30.316
\]

The residual sum of squares is

\[
Res\_SS = TSS_{pooled} - \boldsymbol{\beta}^\prime X^\prime X \boldsymbol{\beta} = 37.544 - 30.316 = 7.228.
\]

Once we have the $Res\_SS$ it is straightforward to construct the OLS standard errors of the regression estimates.  The standard error of each regression estimate is found using the standard approach of multiplying each element along the main diagonal of $(X^\prime X)^{-1}$ by the mean squared error (MSE), where the MSE is $\frac{Res\_SS}{\text{n-regression degrees of freedom}}$, and then taking the square root of each of those products. 

For our example the inverse Gramian matrix is
\[
(X^\prime X)^{-1} =
\begin{bmatrix}
18 & 9 & 6 & 6 \\
9 & 9 & 3 & 3 \\
6 & 3 & 6 & 0 \\
6 & 3 & 0 & 6
\end{bmatrix}^{-1}
=
\begin{bmatrix}
0.2222 & -0.1111 & -0.1667 & -0.1667 \\
-0.1111 & 0.2222 & 0 & 0 \\
-0.1667 & 0 & 0.3333 & 0.1667 \\
-0.1667 & 0 & 0.1667 & 0.3333
\end{bmatrix}
\]

And the MSE $= \frac{7.228}{18-4} = 0.5163$

From which we can construct the standard errors of the estimates as follows 

\[
st\_err(\boldsymbol{\beta}) =
\begin{bmatrix}
st\_err(\beta_0) \\
st\_err(\beta_{\text{treatment\_b}}) \\
st\_err(\beta_{\text{covariate\_2}}) \\
st\_err(\beta_{\text{covariate\_3}})
\end{bmatrix}
=
\begin{bmatrix}
(0.5163 \cdot 0.2222)^{0.5} \\
(0.5163 \cdot 0.2222)^{0.5} \\
(0.5163 \cdot 0.3333)^{0.5} \\
(0.5163 \cdot 0.3333)^{0.5}
\end{bmatrix}
=
\begin{bmatrix}
0.3387 \\
0.3387 \\
0.4148 \\
0.4148
\end{bmatrix}
\]

As well as construct the t-statistics for each estimate as well
\[
t\_stat(\boldsymbol{\beta}) = \frac{\boldsymbol{\beta}}{st\_err(\boldsymbol{\beta})} =
\begin{bmatrix}
t\_stat(\beta_0) \\
t\_stat(\beta_{\text{treatment\_b}}) \\
t\_stat(\beta_{\text{covariate\_2}}) \\
t\_stat(\beta_{\text{covariate\_3}})
\end{bmatrix}
=
\begin{bmatrix}
0.6583 / 0.3387 \\
-0.1188 / 0.3387 \\
0.7211 / 0.4148 \\
1.1159 / 0.4148
\end{bmatrix}
=
\begin{bmatrix}
1.9436 \\
-0.3509 \\
1.7384 \\
2.6900
\end{bmatrix}
\]

Below, as a comparison, we run the regression in Excel on the individual level data.\footnote{To make the TSS and RSS match the Excel results, subtract the $\frac{\sum X^2}{n}$ from each. Since this value cancels out when calculating the residual sum of squares there is no need bother with these subtractions for our use case. Also note that the values in the tables presented are rounded. The regression calculations were performed on data with full precision.} Notice that all the results we calculated above are exactly the same as the results from Excel. 

\begin{table}[ht]
\centering
\caption{Regression Results}
\label{tab:table5}
\begin{tabular}{lccccc}
\toprule
\multicolumn{6}{l}{\textbf{ANOVA}} \\ 
\midrule
 & \textbf{df} & \textbf{SS} & \textbf{MS} & \textbf{F} & \textbf{Significance F} \\
\midrule
Regression & 3 & 3.9060 & 1.3020 & 2.5219 & 0.1000 \\
Residual   & 14 & 7.2279 & 0.5163 &        &        \\
Total      & 17 & 11.1339 &       &        &        \\
\bottomrule
\\

\multicolumn{5}{l}{\textbf{ }}\\ 
\midrule
 & \textbf{Coefficients} & \textbf{Standard Error} & \textbf{t Stat} & \textbf{P-value} \\
\midrule
Intercept  & 0.6583 & 0.3387 & 1.9436 & 0.0723 \\
Assignment & -0.1188 & 0.3387 & -0.3509 & 0.7309 \\
Cov\_2     & 0.7211 & 0.4148 & 1.7384 & 0.1041 \\
Cov\_3     & 1.1159 & 0.4148 & 2.6900 & 0.0176 \\
\bottomrule
\end{tabular}
\end{table}

\section{Use Case: Partial F-Tests for A/B Test Interactions (and Conditional Treatment Effects)}

We can use the dummy variable approach to build a simple, yet efficient approach to flag if independent a/b Tests running contemporaneously might be likely to be interfering with one another (this is a useful approach independent of data minimization). The approach we introduce is the use the f-statistic of the partial f-test of nested models as an omnibus test for possible interactions.  The f-test is constructed by calculating the residual sum of squares for each of the following two models: 

\begin{enumerate}
    \item The main effects regression model (no interactions).
    \item The fully interacted model.
\end{enumerate}
The partial f-statistic for the test is of the following form:

\[
f\_Stat = \frac{\frac{Res\_SS_{\text{main}} - Res\_SS_{\text{full}}}{p}}{\frac{Res\_SS_{\text{full}}}{N-k}}
\]

Where $Res\_SS_{main}$ is the residuals of the main effects model, $Res\_SS_{full}$ is the residual of fully interacted model, p is the difference in the number of parameters (the regression degrees of freedom) of the full model and main effects model, and $k$ is the number of parameters in the full model.  The f-test approach has the useful advantage over the single regression t-test approach in that it provides a single test even in cases where there are more than 2 arms in either of the potentially clashing a/b tests. 

For example, assume that the covariate data in Table~\ref{tab:table2} represents each of three possible treatments from a separate, independently running a/b test.  To run an interaction test, we need to calculate $Res\_SS_{main}$ and $Res\_SS_{full}$ (the $Res\_SS_{main}$ was calculated in the preceding section as 7.228).

To estimate $Res\_SS_{full}$ we extend the design matrix by two dimensions to include the interaction terms between the two tests as Treatment\_b $\cdot$ Cov\_2 and Treatment\_b $\cdot$ Cov\_3.  The Gramian matrix for the full model is from Table~\ref{tab:table2} is

\[
X^\prime X_{\text{full}} =
\begin{bmatrix}
\textit{All Rows} & R4, R5, R6 & R2, R5 & R3, R6 & R5 & R6 \\
R4, R5, R6 & R4, R5, R6 & R5 & R6 & R5 & R6 \\
R2, R5 & R5 & R2, R5 & 0 & R5 & 0 \\
R3, R6 & R6 & 0 & R3, R6 & 0 & R6 \\
R5 & R5 & R5 & 0 & R5 & 0 \\
R6 & R6 & 0 & R6 & 0 & R6
\end{bmatrix}
=
\begin{bmatrix}
\mathbf{18} & \mathbf{9} & \mathbf{6} & \mathbf{6} & 3 & 3 \\
\mathbf{9} & \mathbf{9} & \mathbf{3} & \mathbf{3} & 3 & 3 \\
\mathbf{6} & \mathbf{3} & \mathbf{6} & \mathbf{0} & 3 & 0 \\
\mathbf{6} & \mathbf{3} & \mathbf{0} & \mathbf{6} & 0 & 3 \\
3 & 3 & 3 & 0 & 3 & 0 \\
3 & 3 & 0 & 3 & 0 & 3
\end{bmatrix}
\]

Similarly, we construct $X^\prime y$ by extending the vector to include the conditional sums of $y$, given our two extra interaction terms

\[
X^\prime y_{\text{full}} =
\begin{bmatrix}
\textit{All Rows} \\
R4, R5, R6 \\
R2, R5 \\
R3, R6 \\
R5 \\
R6
\end{bmatrix}
=
\begin{bmatrix}
\mathbf{21.8030} \\
\mathbf{10.3667} \\
\mathbf{7.9204} \\
\mathbf{10.2891} \\
1.7095 \\
7.2345
\end{bmatrix}
\]

Notice that the entries in bold in $X^\prime X_{full}$ are equal to $X^\prime X_{main}$- as it is a submatrix of $X^\prime X_{full}$ - and the bold entries in $X^\prime y_{full}$  are $X^\prime y_{main}$. This fact can help during computation, as  $X^\prime X_{main}$ and $X^\prime y_{main}$ are created in the process of constructing $X^\prime X_{full}$ and $X^\prime y_{full}$. This nesting is also what allows us to use the partial f-test, since all the terms in the simpler model are contained in the full model. 

The regression weights for the full model are calculated as 

\[
\boldsymbol{\beta}_{\text{full}} =
\begin{bmatrix}
\beta_0 \\
\beta_{\text{treatment\_b}} \\
\beta_{\text{covariate\_2}} \\
\beta_{\text{covariate\_3}} \\
\beta_{\text{treatb $\cdot$ cov\_2}} \\
\beta_{\text{treatb $\cdot$ cov\_3}}
\end{bmatrix}
=
\begin{bmatrix}
\mathbf{18} & \mathbf{9} & \mathbf{6} & \mathbf{6} & 3 & 3 \\
\mathbf{9} & \mathbf{9} & \mathbf{3} & \mathbf{3} & 3 & 3 \\
\mathbf{6} & \mathbf{3} & \mathbf{6} & \mathbf{0} & 3 & 0 \\
\mathbf{6} & \mathbf{3} & \mathbf{0} & \mathbf{6} & 0 & 3 \\
3 & 3 & 3 & 0 & 3 & 0 \\
3 & 3 & 0 & 3 & 0 & 3
\end{bmatrix}^{-1}
\begin{bmatrix}
\mathbf{21.8030} \\
\mathbf{10.3667} \\
\mathbf{7.9204} \\
\mathbf{10.2891} \\
1.7095 \\
7.2345
\end{bmatrix}
=
\begin{bmatrix}
0.7236 \\
-0.2494 \\
1.3467 \\
0.2946 \\
-1.2511 \\
1.6427
\end{bmatrix}
\]

From the regression weights we can uncover the regression sum of squares for the full model as

\[
\boldsymbol{\beta}_{\text{f}}^\prime X^\prime X_{\text{f}} \boldsymbol{\beta}_{\text{f}} =
\begin{bmatrix}
0.7236 & -0.2494 & 1.3467 & 0.2946 & -1.2511 & 1.6427
\end{bmatrix}
\begin{bmatrix}
\mathbf{18} & \mathbf{9} & \mathbf{6} & \mathbf{6} & 3 & 3 \\
\mathbf{9} & \mathbf{9} & \mathbf{3} & \mathbf{3} & 3 & 3 \\
\mathbf{6} & \mathbf{3} & \mathbf{6} & \mathbf{0} & 3 & 0 \\
\mathbf{6} & \mathbf{3} & \mathbf{0} & \mathbf{6} & 0 & 3 \\
3 & 3 & 3 & 0 & 3 & 0 \\
3 & 3 & 0 & 3 & 0 & 3
\end{bmatrix}
\begin{bmatrix}
0.7236 \\
-0.2494 \\
1.3467 \\
0.2946 \\
-1.2511 \\
1.6427
\end{bmatrix}
\]

which reduces to
\[
\boldsymbol{\beta}_{\text{f}}^\prime X^\prime X_{\text{f}} \boldsymbol{\beta}_{\text{f}} = 36.634.
\]

Finally, the residual sum of squares for the full model is

\[
Res\_SS_{full} = TSS_{pooled} - \boldsymbol{\beta}_{\text{f}}^\prime X^\prime X_{\text{f}} \boldsymbol{\beta}_{\text{f}} = 37.544 - 36.634 = 0.909
\]

Once we have both the $Res\_SS_{main}$ and the $Res\_SS_{full}$ it is fairly simple to construct the f-statistic of an interaction between the two a/b tests. 

\[
f\text{-statistic} =
\frac{\frac{\text{Res\_SS}_{\text{main}} - \text{Res\_SS}_{\text{full}}}{p}}
{\frac{\text{Res\_SS}_{\text{full}}}{N - k}}
=
\frac{\frac{(7.228 - 0.909)}{(6 - 4)}}
{\frac{0.909}{(18 - 6)}}
= 41.705
\]

Notice that the partial f-test approach does not require constructing the standard errors nor the t-statistics of the regression weights as one would with the t-test of the interaction terms approach. 

As noted, the f-test approach has just one statistic for each combination of a/b tests under consideration. This is useful for a couple of reasons when there are more than two arms in any of the a/b test. Firstly,  there is just one primary statistic to calculate per test combination regardless of number of arms. Secondly, this allows for a cleaner use of any of the family-wise or false discovery rate procedures one is likely to employ when testing over many pairwise checks (of course it is also true that if there will only be 2 arm a/b tests, then there is little practical difference to the f-test approach and the t-test approach.) In addition, if one wanted, after a significant result of the partial f-test, it is easy to construct and inspect each of the t-statistics of each of the individual interaction terms from full regression model. 

The total number of tests of interference/interactions will be C(T,2), where T is the number of concurrent tests. For example, with 100 concurrent tests, without any prior filtering or editorial exclusions there will be up to 4,950 pairwise a/b test interaction evaluations. Under a global null one is very likely to trigger many false positive alerts when using common nominal settings of type 1 error control.  This leads to the importance of using multiple comparison corrections. Which multiple comparison approach to use to mitigate false positives will depend on the cost structure of the confusion matrix. For example, a more conservative family-wise control approach (Bonferroni, Sidak etc) may make more sense if false positives – flag an interaction when there is none - are much more expensive than false negatives – don’t flag when there is an interaction. This is likely to be the case if the review of any flagged pairwise a/b test is a costly endeavor, both in terms of labor and company confidence in the experimentation program’s output.  

This same f-test approach can also be used to ‘discover’ whether or not certain categorical user segments might have heterogenous treatment effects – some users might respond better to one treatment, and others to another treatment.  Rather than cross two separate a/b tests, we instead cross the a/b Test’s treatments with each of the test’s associated categorical user trait/segment data. The rest of the analysis is essentially the same as above.

Often heterogeneity analysis is followed up by a confirmatory a/b test.   Given this protection, surfacing potentially meaningful user segments might have relatively low false-positive error costs, as any targeted action is subject to a follow up test. Therefore, a less conservative false discovery approach, such as Benjamini-Hochberg might be more appropriate in this use case.

One might give some pause that under the privacy approach presented here, it will mean the storing of up to C(T,2) separate data tables – one for each pairwise combination. However, recall that each of these tables are extremely small. As we have seen, for these pairwise test, the number of rows is limited to the cross between the number of treatments in each test and the number of data columns limited to the counts, the sums of each of end points of interest, along with an extra field for the TSS for each end point.  The upper limit will more likely be bound by the reasonableness of the number of statistical comparisons rather than the memory complexity. It is also true that under the micro-data approach one would still need to store $N \cdot T$ records.  

\section{OLS for Low Cardinality Numeric Covariates}

Regression analysis based on k-anonymous aggregated data can be extended beyond dummy coding to ANCOVA problems. Unlike in the dummy coding approach, where just the counts and conditional counts of the dummy variables are used to construct the Gramian matrix, for ANCOVA the squared values of each level in the covariate data are used to weigh the counts in each equivalence class.  The ANCOVA Gramian matrix for equivalence class data is as follows:

\[
X^\prime X =
\begin{bmatrix}
N & \cdots & \sum (\text{Covariate}_{V_{(i)}} \cdot \text{Count}_{(i)}) \\
\vdots & \ddots & \vdots \\
\sum (\text{Covariate}_{V_{(i)}} \cdot \text{Count}_{(i)}) & \cdots & \sum \big((\text{Covariate}_{V_{(i)}})^2 \cdot \text{Count}_{(i)}\big)
\end{bmatrix}
\]

Notice that along the main diagonal, rather than squaring each value and then summing them, $\sum_{i=1}^N x_i^2$  , we instead square the covariate’s value in each row of the equivalence class and then multiply that squared value by the row’s count as follows $\sum ((\text{Covariate}_{\text{value}(i)})^2 \cdot \text{Count}_{(i)})$.

\section{Regression Adjustment Example}
To illustrate, let us consider regression adjustment, a variance reduction technique similar to CUPED \citep{deng2013}, often used for a/b testing when there is ‘useful’ historical or contextual side data. We will follow the ANCOVA2 approach proposed by \citet{lin2013} estimate the regression weights from the equivalence class data. However, rather than use Eicker-White consistent standard errors (for situations when $N_a \neq N_b$), which would require access to the individual level data, we will use an alternative approach suggested by \citet{ding2023}.

Here are the four basic steps for regression adjustment following Lin and Ding:
\begin{enumerate}
    \item Demean the covariate(s) by the pooled covariate(s) mean(s).
    \item Run separate regressions with the covariate(s) for each treatment arm.
    \item Estimate the ATE using the difference of intercept terms from each regression.
    \item Use the residual sum of squares of each regression, following Ding’s suggestion, to construct a Welch-like t-test for the ATE. 
\end{enumerate}

To make the effects of demeaning the data clearer, we will slightly alter Table~\ref{tab:table2} and make the data less perfectly balanced over and within treatments by switching the covariate value of user XXX9 in Table~\ref{tab:table1} from value of ‘3’ to a value of ‘2’. The values in bold in R2 and R3 indicate the values that have been changed. 

\renewcommand{\thetable}{6A}

\begin{table}[ht]
\centering
\caption{Altered Table 2}
\label{tab:table6a}
\begin{tabular}{lccSc}
\toprule
\textbf{Row} & \textbf{Treatment} & \textbf{Covariate} & \textbf{Count} & \textbf{Sum Time on App} \\
\midrule
R1 & A & 1 & 3 & 2.17084932005864  \\
\textbf{R2} & \textbf{A} & \textbf{2} & \B 4 & \textbf{7.09611256990949 } \\
\textbf{R3} & \textbf{A} & \textbf{3} & \B 2 & \textbf{2.16935239750738
} \\
R4 & B & 1 & 3 & 1.42266953605589
 \\
R5 & B & 2 & 3 & 1.70950884373810
 \\
R6 & B & 3 & 3 & 7.23452646813504
 \\
\bottomrule
\end{tabular}
\end{table}

The mean value of the covariate data is 1.944. From each covariate value in Table~\ref{tab:table6a} we subtract 1.944 to de-mean the covariate data as follows in Table~\ref{tab:table6b}

\renewcommand{\thetable}{6B}

\begin{table}[ht]
\centering
\caption{(demean)}
\label{tab:table6b}
\begin{tabular}{lccSc}
\toprule
\textbf{Row} & \textbf{Treatment} & \textbf{Covariate} & \textbf{Count} & \textbf{Sum Time on App} \\
\midrule
R1 & A & {-0.944} & 3 & 2.17084932005864 \\
\textbf{R2} & \B A & \B 0.056 & \B 4 & \B 7.09611256990949 \\
\textbf{R3} & \B A & \B 1.056 & \B 2 & \B 2.16935239750738 \\
R4 & B & {-0.944} & 3 & 1.42266953605589 \\
R5 & B & 0.056 & 3 & 1.70950884373810 \\
R6 & B & 1.056 & 3 & 7.23452646813504 \\
\bottomrule
\end{tabular}
\end{table}

The Gramian matrices for Treatment A and Treatment B from the data in Table~\ref{tab:table6b} are constructed as follows:  

\[
X^\prime X_{\alpha} =
\begin{bmatrix}
N_{\alpha} & \sum (-0.944 \cdot 3) + (0.056 \cdot 4) + (1.056 \cdot 2) \\
\sum (-0.944 \cdot 3) + (0.056 \cdot 4) + (1.056 \cdot 2) & \sum (-0.944^2 \cdot 3) + (0.056^2 \cdot 4) + (1.056^2 \cdot 2)
\end{bmatrix}
\]

\[
=
\begin{bmatrix}
9 & -0.5 \\
-0.5 & 4.9167
\end{bmatrix}
\]

and

\[
X^\prime X_{b} =
\begin{bmatrix}
N_{b} & \sum (-0.944 \cdot 3) + (0.056 \cdot 3) + (1.056 \cdot 3) \\
\sum (-0.944 \cdot 3) + (0.056 \cdot 3) + (1.056 \cdot 3) & \sum (-0.944^2 \cdot 3) + (0.056^2 \cdot 3) + (1.056^2 \cdot 3)
\end{bmatrix}
\]

\[
=
\begin{bmatrix}
9 & 0.5 \\
0.5 & 6.0278
\end{bmatrix}
\]
The $X^\prime y$ vector for those assigned to the ‘A’ and ‘B’ treatments are composed of the sum of y over all users for the intercept term and a weighted sum of y based on the value of the covariate data for the covariate term. More generally

\[
X^\prime y =
\begin{bmatrix}
\sum y_{(i)} \\
\sum \text{Covariate}_{\text{Value}(i)} \cdot y_{(i)}
\end{bmatrix}.
\]

To reiterate, note that when using the aggregate K-anon data, ‘i’ runs from 1 to just the number of equivalence classes for both the $X^\prime X$ and $X^\prime y$ matrices regardless of how many users have been assigned to ‘A’ or ‘B’. So, for our case, ‘i’ runs to just three in each case (if the covariate data had 20 levels, then ‘i’ would run from 1 to 20 etc.).  

\[
X^\prime y_{\alpha} =
\begin{bmatrix}
2.17 + 7.01 + 2.17 \\
(-0.944 \cdot 2.17) + (0.056 \cdot 7.01) + (1.056 \cdot 2.17)
\end{bmatrix}
=
\begin{bmatrix}
11.436 \\
-0.6339
\end{bmatrix}
\]

And 

\[
X^\prime y_{b} =
\begin{bmatrix}
1.42 + 1.71 + 7.23 \\
(-0.944 \cdot 1.42) + (0.056 \cdot 1.71) + (1.056 \cdot 7.23)
\end{bmatrix}
=
\begin{bmatrix}
10.367 \\
6.388
\end{bmatrix}
\]

and the regression adjusted coefficients for treatment A are as follows

\[
\beta_{a} = (X^\prime X)^{-1} X^\prime y =
\begin{bmatrix}
\text{Intercept of A} \\
\text{Covariate}
\end{bmatrix}
=
\begin{bmatrix}
1.285 \\
0.256
\end{bmatrix}
\]

and for treatment B

\[
\beta_{b} = (X^\prime X)^{-1} X^\prime y =
\begin{bmatrix}
\text{Intercept of B} \\
\text{Covariate}
\end{bmatrix}
=
\begin{bmatrix}
1.098 \\
0.969
\end{bmatrix}
\]

The same process used in the ANOVA case is used to calculate the Res\_SS for each regression. However, here we use TSS\_A and TSS\_B respectively to calculate the residual sums of squares for each regression rather than the pooled value. For example, the regression sums of squares for Treatment A’s regression is

\[
Res\_SS_a = TSS_a - \beta^\prime X^\prime X \beta_a = 17.9098-14.8616 =3.0482
\]

And for B

\[
Res\_SS_b = TSS_b - \beta^\prime X^\prime X \beta_b = 19.6336-17.5706 =2.0630
\]

Following \citet{ding2023} we construct a conservative sample variance estimator for the ATE as

\[
\text{Var(SATE)} =
\frac{\text{Res\_SS}_{a}}{n_{a} \cdot (n_{a} - \text{Reg\_DF}_{a})} + \frac{\text{Res\_SS}_{b}}{n_{b} \cdot (n_{b} - \text{Reg\_DF}_{b})} = \frac{3.0482}{9 \cdot (9 - 2)} + \frac{2.0630}{9 \cdot (9 - 2)} = 0.08113
\]

The t-stat for our ATE test is

\[
t_{\text{stat}} = 
\frac{\text{ATE}}{\sqrt{\text{Var(SATE)}}} = 
\frac{(1.098 - 1.285)}{\sqrt{0.08113}} = 
\frac{-0.1871}{0.2848} = 
-0.6568
\]

One can also estimate the Population ATE (PATE) variance as well. Following \citet{imbens2009}, the third term for the normalized variance is given by

\[
V_{\tau} = 
\frac{
\left[\left(\sum x_{\text{demean},a}^2 + \sum x_{\text{demean},b}^2\right) \cdot \left(\beta_{\text{Covariate},b} - \beta_{\text{Covariate},a}\right)\right]^2
}{
N \cdot (N - 1)
}
=
\frac{
\left[(10.9444) \cdot (0.969 - 0.256)^2\right]
}{
18 \cdot (17)
}
\]

\[
V_{\tau} =
\frac{(10.9444) \cdot (0.502727479)}{306}
= 0.01798
\]

Adding $V_\tau$, the population adjustment, to the Var(SATE), the sample variance (Neyman), gives us the variance of the population ATE as follows:

\[
\text{Var(PATE)} = \text{Var(SATE)} + V_{\tau} = 0.08113 + 0.01798 = 0.09911.
\]

The t-stat using the PATE is 

\[
t_{\text{stat}} =
\frac{\text{ATE}}{\sqrt{\text{Var(PATE)}}} =
\frac{(1.098 - 1.285)}{\sqrt{0.09911}} =
\frac{-0.1871}{0.31482} =
-0.5943.
\]

\section{Comparison with OLS on Micro Data}

The following is the Excel regression output of the fully interacted model (ANCOVA) using the pooled OLS standard errors using the individual level data for the SATE. 

\renewcommand{\thetable}{7}
\begin{table}[H]
\centering
\caption{Regression Output for ANCOVA Using Pooled OLS Standard Errors for Individual-Level SATE Data}
\label{tab:table7}
\begin{tabular}{l *{3}{S[table-format=1.4]}}
\toprule
& \textbf{Coefficients} & \textbf{Standard Error} & \textbf{t-Stat} \\
\midrule
Intercept  & 1.2851 & 0.2020 & 6.3627 \\
\textbf{ATE} & \textbf{-0.1871} & \textbf{0.2856} & \textbf{-0.6551} \\
Covariate  & 0.2596 & 0.2733 & 0.9500 \\
Interaction & 0.7090 & 0.3681 & 1.9260 \\
\bottomrule
\end{tabular}
\end{table}

Notice that, as expected, the estimate of the ATE is the same as the separate regression model per arm approach using the K-anon aggregated data. However, there is a very slight difference between standard error and t-statistic for the ATE in the standard pooled Excel results and the results we calculated by ‘hand’ for the SATE. This difference is not due to use of the K-anon data, but rather that the two approaches estimated the standard error differently. The Excel output uses the OLS standard error estimates, rather than the approach suggested by \citet{ding2023} in order to account for heteroskedastic errors.

\section{Conclusion}
Beyond the many contexts in regulated industries, such as Health (HIPAA), and Finance, where access to microdata is strictly regulated, the incorporation of privacy by design directly into privacy regulations means that regardless of industry, companies should, at a minimum, be intentional about the collection of the marginal bit of data. In many cases individual level data is not needed for many standard analyses that are part of the experimentation tool kit - specifically the methods that are based on OLS. Collection and storage of data in equivalence classes, while reducing optionality, has the following positive properties:
\begin{enumerate}
    \item The data is transparent and easily allows of use of K-Anonymity as a measure of data minimization and privacy risk.
    \item Is efficient with respect to computation and memory use.
\end{enumerate}
This is especially true when using the local privacy collection approach, since the equivalence class data are collected without ever storing individual level data.  You can’t lose what you never had. 

Of course, there are many situations where individual level data needs to be collected in order to complete certain required tasks. However, for privacy reasons, this data may not be appropriate to share, even internally.  The k-anonymous, equivalence class approach, might provide a way to unlock some of this sensitive data by eliminating the need to share copies of sensitive data. By k-anonymizing the data, one can decrease the risk of exposure.  In many cases, companies are able to internally share data aggregated as long as the expected marginal utility of the granularity/precision balances the assessed risk. This can open up the use of valuable data that previously would not been accessible. 

\section{Acknowledgments}

I thank Winston Lin for his help, particularly in pointing me to both Woodford and Imbens and to Ding.  I thank Andrew Gershoff, Tim Wilson, Dan Lass, and Tamer Abdelgawad for their feedback on various drafts.

\bibliographystyle{unsrtnat}
\bibliography{references}  

\end{document}